\documentclass[10pt,a4paper,english,pra, twocolumn]{revtex4-1}
\usepackage[T1]{fontenc}
\usepackage[latin9]{inputenc}
\usepackage{color}
\definecolor{page_backgroundcolor}{rgb}{1, 1, 1}
\pagecolor{page_backgroundcolor}
\usepackage{booktabs}
\usepackage{amsmath}
\usepackage{amssymb}
\usepackage{graphicx}
\usepackage{wasysym}

\makeatletter

\providecommand{\tabularnewline}{\\}

\usepackage[colorlinks = true,
            linkcolor =red,
            urlcolor  =magenta,
            citecolor = blue,
            anchorcolor = blue]{hyperref}
\begingroup

\endgroup

\makeatother

\usepackage{babel}
\begin{document}
\title{Generalized Heralded Generation of Non-Gaussian States Using an Optical
Parametric Amplifier }
\author{Xiao-Xi Yao, Bo Zhang}
\author{Yusuf Turek}
\email{Corresponding author: yusufu1984@hotmail.com}

\affiliation{School of Physics,Liaoning University,Shenyang,Liaoning 110036,China}
\begin{abstract}
The heralded optical parametric amplifier (OPA) has emerged as a promising
tool for quantum state engineering. However, its potential has so
far been limited to coherent state inputs. Here, we introduce a generalized
heralded OPA protocol that unlocks a vastly expanded class of quantum
phenomena by accepting arbitrary nonclassical inputs. With a squeezed
vacuum input, the setup functions as an integrated two-photon subtractor,
non-deterministically generating high-fidelity, larger-amplitude squeezed
Schrödinger cat (SC) states---an operation previously requiring complex,
discrete setups. Furthermore, when fed a small-amplitude SC state,
the protocol acts as a non-Gaussianity amplifier, distilling it into
high-purity approximations of key quantum resources like specific
photon-number superpositions. This work transforms the OPA from a
specialized source into a versatile platform for advanced quantum
state engineering. The integrated setup enables the generation of
a wide array of non-Gaussian states, providing a unified approach
to quantum state synthesis.
\end{abstract}
\maketitle

\section{\label{sec:1}Introduction}

Non-Gaussian (nG) states \citep{PRXQuantum.2.030204,PRXQuantum.1.020305}
are indispensable resources in quantum information science. By going
beyond the limitations of the Gaussian framework, they are crucial
for achieving a continuous-variable (CV) quantum computational advantage
\citep{PhysRevLett.82.1784,PhysRevA.64.012310,PhysRevLett.97.110501,PhysRevA.93.022301,PhysRevLett.109.230503},
for enabling fault-tolerant quantum computation with encoded qubits
\citep{PhysRevX.8.021054}, and for enhancing quantum metrology \citep{PhysRevLett.104.103602,PhysRevA.78.063828}.
nG states are typically characterized by the negativity of their Wigner
functions and the presence of higher-order quantum correlations. Among
the most extensively studied nG states are photon-added coherent states
\citep{PhysRevA.43.492}, squeezed Fock states \citep{PhysRevA.40.2494},
and Schrödinger cat (SC) states \citep{2019Duan}. However, the generation
of high-quality nG states remains experimentally challenging due to
practical limitations of existing methods.

Traditional approaches for producing nG states generally fall into
two categories: nonlinear optical processes \citep{PhysRevLett.57.13,PhysRevA.55.2478}
and conditional measurements such as photon addition/subtraction \citep{2007S},
photon catalysis \citep{PhysRevLett.88.250401} and weak value amplification
technique \citep{PhysRevA.105.022608,RN15}. While nonlinear optical
schemes can, in principle, deterministically generate nG states, their
implementation is severely constrained by the weakness of optical
nonlinearities. In addition, such schemes often suffer from poor scalability
and strong susceptibility to losses, which limit their practical feasibility.
In recent years, substantial efforts have been devoted to generating
nG states from Gaussian states via conditional measurements. The key
idea is to realize effective nG operations---such as photon addition
or subtraction \citep{2004S,PhysRevLett.92.153601,Wakui:07,PhysRevLett.97.083604,PhysRevA.82.063833,PhysRevA.110.023703}
by beam splitters. For example, subtracting a single photon from a
squeezed vacuum (SV) state can produce an odd SC state \citep{2006};
however, the resulting SC states usually possess small coherent amplitudes
($\alpha<1.2$) and therefore exhibit only weak non-Gaussianity. In
many quantum technologies, stronger nG resources are essential to
achieve genuine performance advantages \citep{PhysRevA.64.022313,PhysRevA.64.012310,RevModPhys.77.513,PhysRevA.99.032327,Walschaers_2023}.
Another well-known example is the generation of SC states with arbitrary
amplitudes ($\alpha=\sqrt{n}$ for $n\ge3$) by performing conditional
measurements on Fock states $\vert n\rangle$ \citep{ourjoumtsev2007generation}
and two-mode Gaussian states \citep{PhysRevA.103.013710}. However,
producing high-purity $n$ photon number states and their effective
detection remains technically demanding, which makes the practical
implementation of this approach challenging. Additionally, most existing
models are restricted to generating only a specific type of nG state,
lacking the flexibility and universality needed for a scalable quantum
technology.

The preparation of nonclassical light relies predominantly on nonlinear
media, with second-and third-order nonlinearities being particularly
important. Among second-order nonlinear devices, the optical parametric
amplifier (OPA) stands as a fundamental component for generating a
variety of quantum states. Its versatility allows for the effective
generation of both two-mode and single-mode SV states under non-degenerate
and degenerate operational phase matching conditions, respectively.
A recent seminal study by Shringarpure and Franson \citep{PhysRevA.100.043802}
further demonstrated the OPA's power in a heralded configuration.
In their scheme, injecting a coherent state and a single photon into
the signal and idler modes, respectively, creates entanglement between
them. The subsequent detection of a single photon in the output idler
mode then heralds the generation of useful nG states at the signal
output, such as single-photon-added coherent (SPAC) and displaced
single-photon states. While seminal, this scheme was confined to a
single class of input states (coherent states), leaving open a fundamental
question: does the true power of the heralded OPA lie in its ability
to process already non-classical states, thereby acting as a nonlinear
quantum processor rather than just a state generator?

In this work, we affirmatively answer this question by generalizing
the heralded OPA protocol to accommodate arbitrary non-classical signal
states. We specifically demonstrate that using a SV state as the input
enables the deterministic engineering of a high-fidelity, larger-amplitude
squeezed SC state with a coherent amplitude of approximately $\alpha\approx1.4$---an
operation functionally equivalent to an integrated two-photon subtraction.
Furthermore, we show that when the input is already a nG state, such
as a small-amplitude SC state, our protocol acts as a powerful amplifier
of non-Gaussianity. By tuning the amplitude gain $g$, we can significantly
enhance the Wigner negativity of both even and odd SC states, distilling
them into high-purity approximations of key quantum resources. The
significance of our generalized system is threefold: it is (1) scalable,
as it leverages integrated optics; (2) practical, offering high success
probabilities commensurate with other heralding schemes; and (3) versatile,
capable of generating a wide class of high-quality nG states from
a single setup. This establishes our generalized heralded OPA protocol
as a promising and unified platform for advancing quantum optics and
quantum information processing.

The paper is organized as follows. In Sec. \ref{sec:2}, we review
the protocol for nG state generation based on the OPA. In Sec. \ref{sec:3}
and \ref{sec:4}, we discuss the cases of a SV input and a small-amplitude
SC state input, respectively. In these sections, we analyze the output
states through their Wigner functions and Wigner-negative volumes
to quantitatively characterize their nG features. In Sec. \ref{sec:5},
we briefly discuss the effects of photon loss on the nonclassicality
of generated states. Finally, we present a discussion and summarize
our work in Sec. \ref{sec:6}.

\section{\label{sec:2} Model setup }

Our work is motivated by Ref. \citep{PhysRevA.100.043802}; a schematic
of our setup is shown in Fig. 1. In this scheme the action of OPA
plays an essential role. We therefore begin with a brief introduction
to the OPA's function. The Hamiltonian of an OPA can be described
by the three-wave mixing device \citep{PhysRevA.31.3093,PhysRevA.86.063802}
\begin{equation}
H=\hbar\omega_{s}a^{\dagger}a+\hbar\omega_{i}b^{\dagger}b+\hbar\omega_{p}c^{\dagger}c+i\hbar\chi^{(2)}\left(abc^{\dagger}-a^{\dagger}b^{\dagger}c\right),\label{eq:1}
\end{equation}
where $a$, $b$, and $c$ are the annihilation operators of the signal,
idler, and pump with frequencies $\omega_{s}$, $\omega_{i}$, and
$\omega_{p}$, respectively, and $\chi^{(2)}$ is the coupling strength,
determined by the second-order nonlinear susceptibility of the BBO
crystal. Assuming the pump field is a high-amplitude coherent state
$\vert\gamma e^{-i\omega_{p}}\rangle$ and can therefore be treated
classically, we rewrite the Hamiltonian in the interaction picture
with $\omega_{p}=\omega_{s}+\omega_{i}$ as 
\begin{equation}
H_{I}=i\hbar\left(\xi^{\ast}ab-\xi a^{\dagger}b^{\dagger}\right),\label{eq:2-1}
\end{equation}
where $\xi=\gamma\chi^{(2)}.$ The unitary operator corresponding
to this interaction Hamiltonian reads 
\begin{equation}
U_{I}(t)=e^{-iH_{I}t/\hbar}=\exp\left[\tau^{\ast}ab-\tau a^{\dagger}b^{\dagger}\right]\equiv S(\tau),\ \ \tau=\varrho e^{i\delta}.\label{eq:3-1}
\end{equation}
 This is a two-mode squeezing operator with a complex squeezing parameter
$\tau$. By using the OPA, in the Heisenberg picture, the annihilation
operator of the signal mode undergoes the transformation 
\begin{align}
a_{out} & =S^{\dagger}(\tau)aS(\tau)=ga-b^{\dagger}e^{i\delta}\sqrt{g^{2}-1}\nonumber \\
 & =ga+L^{\dagger},\label{eq:4-2}
\end{align}
where $g=\cosh\varrho$ and $L=-b^{\dagger}e^{i\delta}\sinh\varrho$
is the noise operator and one usually assumes that $\langle L^{\dagger}\rangle=0$.
Eq. (\ref{eq:4-2}) represents the physical requirement of an optical
amplifier in terms of input-optput relation \citep{PhysRev.128.2407}.
By passing in this kind of nonlinear medium, the mean photon number
of input signal beam increases with increasing the parameter $\varrho$,
and we refer to $g$ as the amplitude gain. This is a key reason why
the second-order nonlinear medium described by the interaction Hamiltonian
$H_{I}$ is termed an OPA . 
\begin{figure}
\includegraphics[width=8cm]{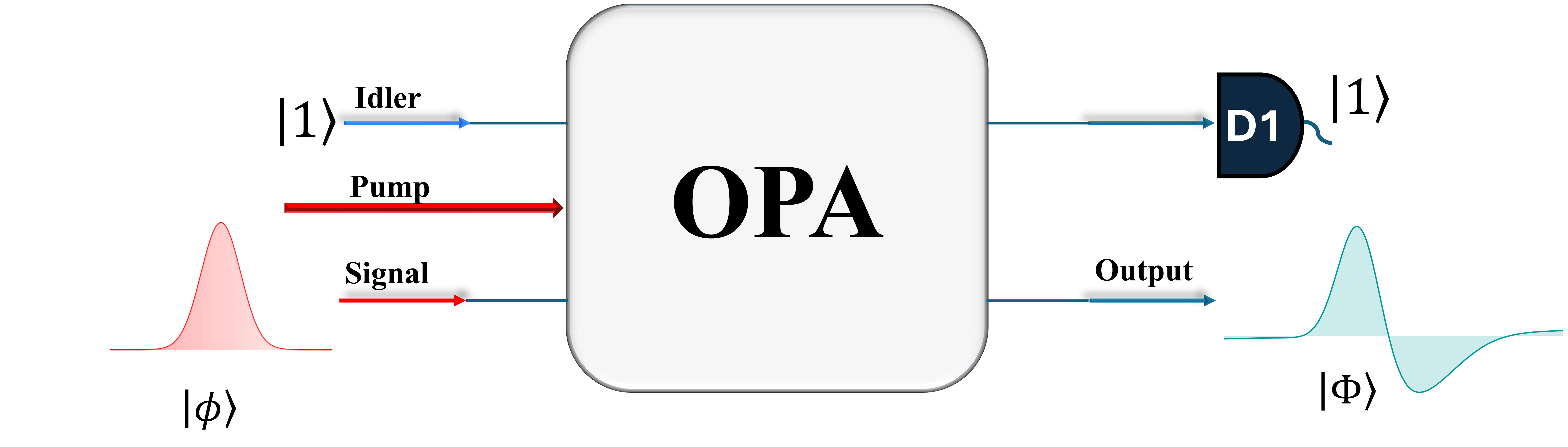}

\caption{\label{fig:1}Schematic of non-Gaussian state generation using an
OPA. Specific states $\vert\phi\rangle$-- such as squeezed vacuum
and small-amplitude Schrödinger cat (SC) states - are injected into
the signal mode, while a single photon is input into the idler mode
of the OPA. A single-photon detection at the idler output port heralds
the generation of a non-Gaussian quantum state $\vert\Phi\rangle$
at the signal output. The characteristics of the output state can
be controlled by adjusting the gain of the amplifier.}
\end{figure}
If the signal and idler modes are initially in the vacuum state, $\vert0\rangle_{a}\vert0\rangle_{b}$,
the OPA transforms them into a two-mode squeezed vacuum (TMSV) state:
\begin{align}
\vert\tau\rangle & =S(\tau)\vert0,0\rangle=\exp\left[\tau^{\ast}ab-\tau a^{\dagger}b^{\dagger}\right]\vert0\rangle_{a}\vert0\rangle_{b}\nonumber \\
 & =\frac{1}{\cosh\varrho}\sum_{n=0}^{\infty}(-e^{i\delta}\tanh\varrho)^{n}\vert n,n\rangle.\label{eq:4-1}
\end{align}
 This state is not a product of two single-mode SV states but is instead
an entangled state exhibiting strong correlations between the two
modes. Note that for a TMSV state, the reduced density matrix of either
mode is a classical state---specifically, a thermal state: 
\begin{equation}
\rho_{th}=(1-\lambda_{r})\sum_{n=0}^{\infty}\lambda_{r}^{n}\vert n\rangle\langle n\vert=\sum_{n=0}^{\infty}\frac{\langle n\rangle_{th}^{n}}{\left(1+\langle n\rangle_{th}\right)^{n+1}}\vert n\rangle\langle n\vert.\label{eq:5-1}
\end{equation}
Here, $\lambda_{r}=\tanh r$ and $\langle n\rangle_{th}=Tr\left(a^{\dagger}a\rho_{th}\right)=Tr\left(b^{\dagger}b\rho_{th}\right)$
represents the average number of excitations. The thermal state, which
describes each mode of the TMSV state, has a Gaussian profile in phase
space. Furthermore, in the degenerate case (i.e., $a=b$) the two-mode
squeezed operator $S(\tau)$ becomes the conventional squeezed operator
as $\exp[\tau^{\ast}a^{2}-\tau a^{\dagger2}]$. In this case, if the
initial signal state is prepared in vacuum and coherent states, the
output state gives SV and squeezed coherent states, respectively.

However, in Ref. \citep{PhysRevA.100.043802} the authors introduced
the nG state generation proposal based on OPA with conditional measurement
carried out on the idler mode. As shown in schematic, Fig. \ref{fig:1},
if the signal mode is prepared in a coherent state and idler mode
contains single-photon, then some typical nG states can be generated
at signal output if a single-photon heralded in output idler mode. 

If we consider the general case, the conditional state generation
process via OPA can be described as follows: 

(i) The signal mode input is prepared in single mode radiation field
state $\vert\phi\rangle$ while the idler mode contains single-photon. 

(ii) The initial composite state $\vert\phi\rangle_{s}\vert1\rangle_{i}$
passes through OPA and is transformed into $S\vert\phi\rangle_{s}\vert1\rangle_{i}$.
Here, $S$ denotes the factored form of two-mode squeezed operator
$S(\tau)$, given by 
\begin{equation}
S=\frac{1}{g}e^{-Ga^{\dagger}b^{\dagger}}e^{-(a^{\dagger}a+b^{\dagger}b)\ln g}e^{Gab},\label{eq7}
\end{equation}
 where $G=\sqrt{g^{2}-1}/g$. 

(iii) A conditional measurement, specifically the detection of a single-photon
in the output idler mode, is represented by the projector $\ensuremath{\Pi=|1\rangle_{i}\langle1|}$.
A successful detection heralds the preparation of the output signal
state:

\begin{equation}
\left|\psi_{out}\right\rangle =\mathcal{N}{}_{i}\langle1|\Pi S|\phi\rangle_{s}|1\rangle_{i}=\frac{_{i}\langle1|\Pi S|\phi\rangle_{s}|1\rangle_{i}}{\sqrt{P_{f}}},\label{eq:2}
\end{equation}
The factor $\mathcal{N}$ is the normalization constant, and $P_{f}=\vert\mathcal{N}\vert^{-2}$
represents the success probability for preparing the signal-mode state.
This probability quantifies the likelihood that, in a given experimental
trial, a single photon emerges in the output idler mode to herald
the successful generation of the target nG state. In a practical implementation,
the success probability is reduced by the finite efficiency $\eta$
of the single-photon detector. Thus, the effective single-shot success
probability is given by $P=\eta P_{f}$. With current high-efficiency
superconducting single-photon detectors ($\eta$ between $0.84$ and
$0.99$) \citep{Korzh2020np,chang2021detecting,11097370}, the detector
loss is already small. Recent advances with superconducting transition-edge
sensors (TESs) have further pushed the efficiency frontier, reporting
values exceeding $95\%$ ($\eta=0.95$) in the telecom band while
offering photon-number resolution capabilities \citep{10.1063/5.0149478,Li:25}.
For our protocol, which relies on heralding a single-photon outcome,
such high efficiency makes detector loss negligible for the fidelity
of the generated state. This is because the expected photon number
is low. Moreover, significant infidelity would require correlated
errors, which are highly improbable with these efficient detectors.
Therefore, for the purpose of analyzing the state fidelity, we can
safely approximate $P\approx P_{f}$. The intrinsic success probabilities
$P_{f}$, which depend on the specific input state $\vert\phi\rangle_{s}$
and the OPA gain $g$, will be evaluated in the following sections. 

Note that in this protocol, the initial signal state $\vert\phi\rangle$
is arbitrary. From this perspective, the state generation protocol
can be applied to any class of signal states. Following the derivation
of Ref. \citep{PhysRevA.100.043802}, for a coherent state input $\vert\phi\rangle=\vert\alpha\rangle_{s}=D(\alpha)\vert0\rangle$,
the unnormalized output state of the signal mode is given by:

\begin{equation}
\left|\psi_{\text{out }}\right\rangle \propto\left(\frac{1}{g^{2}}-\frac{G^{2}}{g}\alpha a^{\dagger}\right)|\alpha/g\rangle_{s}.\label{eq:3}
\end{equation}
Here, $\vert\alpha/g\rangle_{s}$ represents an attenuated version
of the initial signal state. Equation (\ref{eq:3}) is the central
result of Ref. \citep{PhysRevA.100.043802}, showing that by tuning
the amplitude gain $g$, one can generate specific nG states such
as single-photon-added coherent (SPAC) states and displaced single
photon number (DSPN) states with high fidelity. 

As previously noted, this protocol generalizes to non-coherent inputs.
In the following sections, we demonstrate that using a SV state or
a small-amplitude SC state as the signal input enables the generation
of other useful nG states with extremely high fidelity.

\section{\label{sec:3}Squeezed Vacuum Signal Input }

We now consider the case where the signal input is SV state, $\vert\phi\rangle_{a}=\vert\xi\rangle=S(\xi)\vert0\rangle$.
Here, $S(\xi)=exp[\frac{1}{2}(\xi^{\ast}a^{2}-\xi a^{\dagger2})]$
is the single-mode squeezing operator with a complex squeezing parameter
$\xi=re^{i\theta}$. The SV state $\vert\xi\rangle$ can be written
in terms of Fock states as 
\begin{equation}
\vert\xi\rangle=\frac{1}{\sqrt{\cosh r}}\sum_{n-0}^{\infty}e^{in\theta}(\tanh r)^{n}\frac{\sqrt{(2n)!}}{n!2^{n}}\vert2n\rangle.\label{eq:4}
\end{equation}
Following the protocol described in Sec. \ref{sec:2}, the unnormalized
output state of the signal mode is determined by the conditional expression:

\begin{align}
|\psi\rangle & =\Pi S|\eta\rangle_{a}|1\rangle_{b}\nonumber \\
 & =\Pi\frac{1}{g}e^{-Ga^{\dagger}b^{\dagger}}e^{-(a^{\dagger}a+b^{\dagger}b)\ln g}e^{Gab}|\xi\rangle|1\rangle_{b}\label{eq:5}
\end{align}
To derive an explicit expression, we begin by applying the operator
$e^{Gab}$ to the initial state: 

\begin{align}
\vert\psi^{\prime}\rangle=e^{Gab}|\phi\rangle_{a}|1\rangle_{b} & =\left(1+Gab+\cdots\right)|\phi\rangle_{a}|1\rangle_{b}\nonumber \\
 & =|\phi\rangle_{a}|1\rangle_{b}+Gab|\phi\rangle_{a}|1\rangle_{b}\nonumber \\
 & =|\phi\rangle_{a}|1\rangle_{b}-Ge^{i\theta}\sinh rS(\xi)|1\rangle_{a}|0\rangle_{b}.\label{eq:12}
\end{align}
Next, we apply the operator $g^{-(a^{\dagger}a+b^{\dagger}b)}$ to
$\vert\psi^{\prime}\rangle$: 
\begin{align}
\vert\psi^{\prime\prime}\rangle & =g^{-(a^{\dagger}a+b^{\dagger}b)}\vert\psi^{\prime}\rangle\nonumber \\
 & =\frac{1}{g}S(\xi)e^{-B\ln g}|0\rangle_{a}\vert1\rangle_{b}-Ge^{i\theta}\sinh rS(\xi)e^{-B\ln g}|1\rangle_{a}\vert0\rangle_{b},\label{eq:7}
\end{align}
where the operator $B$ is defined as 
\begin{equation}
B=a^{\dagger}a\cosh^{2}r+aa^{\dagger}\sinh^{2}r-\frac{e^{i\theta}}{2}a^{\dagger2}\sinh2r-\frac{e^{-i\theta}}{2}a^{2}\sinh2r.\label{eq:8}
\end{equation}
 We then apply the projector and the remaining exponential. The action
of the projector on the first exponential term is:
\begin{align}
\Pi e^{-Ga^{\dagger}b^{\dagger}} & =|1\rangle_{bb}\langle1|\left(1-Ga^{\dagger}b^{\dagger}+\cdots\right)\nonumber \\
 & =|1\rangle_{bb}\langle1|-Ga^{\dagger}|1\rangle_{bb}\langle0|.\label{eq:10}
\end{align}
 Substituting the expressions, Eqs. (\ref{eq:7}) and (\ref{eq:10})
into the Eq. (\ref{eq:5}) yields the final unnormalized output state
for the signal mode:
\begin{align}
|\Psi\rangle & =\frac{1}{g}\left[\frac{1}{g}S(\xi)e^{-B\ln g}|0\rangle_{a}+G^{2}e^{i\theta}\sinh ra^{\dagger}S(\xi)e^{-B\ln g}|1\rangle_{a}\right].\label{eq:16}
\end{align}
To simplify this expression, we must evaluate the action of $e^{-B\ln g}$
on the Fock states $\vert0\rangle_{a}$ and $\vert1\rangle_{a}$.
We recognize that the operator $B$ is an element of the $SU(1,1)$
Lie algebra. This allows us to factorize the exponential $e^{-B\ln g}$
into a disentangled form, which translates to a known sequence of
squeezing and scaling operations on the vacuum state. To proceed,
we define the generators

\begin{align}
K_{z} & =\frac{1}{2}(a^{\dagger}a+\frac{1}{2}),\\
K_{+} & =\frac{a^{\dagger2}}{2},\ \ K_{-}=\frac{a^{2}}{2},
\end{align}
 which satisfy the $SU(1,1)$ commutation relations \citep{puri2001mathematical}
\begin{equation}
[K_{z},K_{\pm}]=\pm K_{\pm},[K_{+},K_{-}]=-2K_{z}.
\end{equation}
 Hence, we can write $e^{-B\ln g}$ as 
\begin{align}
A & =e^{-B\ln g}\nonumber \\
 & =\lambda exp\left[\frac{\phi_{+}\left(\theta^{\prime}\right)}{2}a^{\dag2}\right]exp\left[\frac{\phi_{z}\left(\theta^{\prime}\right)}{2}a^{\dag}a\right]exp\left[\frac{\phi_{-}\left(\theta^{\prime}\right)}{2}a^{2}\right],\label{eq:14}
\end{align}
where $\theta^{\prime}=-\ln g$, $\alpha_{\pm}=-e^{\pm i\theta}\sinh2r$,
$\alpha_{z}=2\cosh2r$, $\lambda=e^{\frac{1}{2}\left[\ln g+\phi_{z}(\theta^{\prime})\right]}$,
and the coefficients $\phi_{\pm}$, $\phi_{z}$ are given by 

\begin{subequations}

\begin{align}
\varGamma & =\sqrt{\frac{1}{4}\alpha_{z}^{2}-\alpha_{+}\alpha_{-}},\\
\phi_{\pm}\left(\theta^{\prime}\right) & =\frac{\alpha_{\pm}}{\varGamma}\frac{\sinh\varGamma\theta^{\prime}}{\cosh(\varGamma\theta^{\prime})-\alpha_{z}\sinh(\varGamma\theta^{\prime})/2\varGamma},\\
\phi_{z}\left(\theta^{\prime}\right) & =-2\ln\left[\cosh(\varGamma\theta^{\prime})-\frac{\alpha_{z}}{2\varGamma}\sinh(\varGamma\theta^{\prime})\right].
\end{align}
 \end{subequations}

Applying this decomposition to the vacuum state is now simple, as
the rightmost exponential annihilates it: 
\begin{align}
e^{-B\ln g}|0\rangle_{a} & =exp\left[\frac{\phi_{+}\left(\theta^{\prime}\right)}{2}a^{\dag2}\right]\vert0\rangle_{a}\nonumber \\
 & =\lambda\sqrt{\cosh r^{\prime}}S(\xi^{\prime})\vert0\rangle_{a},\label{eq:18}
\end{align}
where $\xi^{\prime}=r^{\prime}e^{i\varphi}$ and the real new squeezing
parameter $r^{\prime}$ and phase $\varphi$ are determined by $\phi_{+}\left(\theta^{\prime}\right)=-e^{i\varphi}\tanh r^{\prime}$.
This shows that the net effect on the vacuum is, up to a normalization,
simply a squeezing operation. A similar, though more involved, calculation
for the $\vert1\rangle_{a}$ state yields
\begin{align}
e^{-B\ln g}|1\rangle_{a} & =\lambda e^{\frac{\phi_{z}\left(\theta^{\prime}\right)}{2}}exp\left[\frac{\phi_{+}\left(\theta^{\prime}\right)}{2}a^{\dag2}\right]\vert1\rangle_{a}\nonumber \\
 & =\lambda e^{\frac{\phi_{z}\left(\theta^{\prime}\right)}{2}}(\cosh r^{\prime})^{\frac{3}{2}}S(\xi^{\prime})\vert1\rangle_{a}.\label{eq:19}
\end{align}
 In the above derivations, we used the Fock basis representation of
the SV state, given in Eq. (\ref{eq:4}. Substituting these results
back into the expression for $\vert\Psi\rangle$, the output state
can be simplified to a compact form:
\begin{equation}
\vert\Psi^{\prime}\rangle=\mathcal{N^{\prime\prime}}S(\xi^{\prime\prime})\left[c_{0}\vert0\rangle+e^{2i\varTheta(\eta,\xi)}c_{2}\vert2\rangle\right],\label{eq:26}
\end{equation}
where $\mathcal{N^{\prime\prime}}=1/\sqrt{\vert c_{0}\vert^{2}+\vert c_{2}\vert^{2}}$
is the normalization constant and 
\begin{align}
c_{0} & =\frac{\lambda\sqrt{\cosh r^{\prime}}}{g^{2}}-\frac{\lambda G^{2}}{g}e^{i\theta}\sinh re^{\frac{\phi_{z}\left(\theta^{\prime}\right)}{2}}(\cosh r^{\prime})^{\frac{3}{2}}\alpha_{1},\\
c_{2} & =\sqrt{2}\frac{\lambda G^{2}}{g}\sinh re^{\frac{\phi_{z}\left(\theta^{\prime}\right)}{2}}(\cosh r^{\prime})^{\frac{3}{2}}\alpha_{2}
\end{align}
with $\alpha_{1}=\sinh r^{\prime}\cosh r+\cosh r^{\prime}\sinh r$,
$\alpha_{2}=\cosh r^{\prime}\cosh r+\sinh r^{\prime}\sinh r$, and
$\xi^{\prime\prime}=r^{\prime\prime}e^{i\varphi^{\prime\prime}}$
defines the new squeezing parameter $r^{\prime\prime}$ and phase
$\varphi^{\prime\prime}$ for the output state which determined by

\begin{equation}
\tanh r^{\prime\prime}e^{i\varphi^{\prime\prime}}=\frac{\zeta_{1}+\zeta_{2}}{1+2\zeta_{1}\zeta_{2}^{*}}
\end{equation}
with $\zeta_{1}=\tanh re^{i\theta},\zeta_{2}=\tanh r^{\prime}e^{i\varphi}$,
and the phase factor $\varTheta(\eta,\xi)$ determined with
\begin{equation}
\varTheta(\eta,\xi)=\frac{1}{i}\ln\left(\frac{1+\zeta_{1}\zeta_{2}^{*}}{1+\zeta_{1}^{*}\zeta_{2}}\right).
\end{equation}
For the parameters ($g$,$r$,$\eta$,$\theta$) considered herein,
the success probability for preparing the state lies in the range
of $10^{-4}$ to $10^{-2}$. This probability decreases with increasing
gain $g$, as it is determined by $\eta|\mathcal{N^{\prime\prime}}|^{-2}$.
In contrast, the quality (fidelity) of the output state saturates
with increasing $g$ . This trade-off implies that by choosing $g$
appropriately, a high-quality nG state $\vert\Psi^{\prime}\rangle$
can be generated with a practically acceptable success probability,
thereby keeping the experimental overhead low.

This result indicates that the output state belongs to a family of
squeezed superpositions of the zero- and two-photon Fock states. The
coefficients of the superposition can be controlled by adjusting the
initial squeezing parameter $r$ and amplitude gain $g$. As shown
in a recent studies such states are important resources for the efficient
generation of high-fidelity, large-amplitude ($\alpha\ge2$) SC states
\citep{2024catstate} and quantum error correction and fault-tolerant
quantum computation tasks \citep{RN14}. Notably, this process is
functionally equivalent to performing two-photon subtraction ($a^{2}$)
from the original SV state, or to a sequence of single-photon-subtracted-and-added
($a^{\dagger}a$): 

\begin{align}
a^{2}\vert r\rangle & \propto S(r)\left[\vert0\rangle-\sqrt{2}\tanh r\vert2\rangle\right],\label{eq:13}\\
a^{\dagger}a\vert r\rangle & \propto S(r)\left[\vert0\rangle-\sqrt{2}\left(\tanh r\right)^{-1}\vert2\rangle\right],\label{eq:31-1}
\end{align}
where $\vert r\rangle=S(r)\vert0\rangle$ with real squeezing parameter
$r$. The equivalence between these two operations can be verified
through the relation 
\begin{align}
a^{\dagger2}\vert r\rangle & =-\coth r\ a^{\dagger}a\vert r\rangle,\label{eq:36}\\
a^{2}\vert r\rangle & =-\tanh r\:aa^{\dagger}\vert r\rangle.\label{eq:eq34}
\end{align}
 
\begin{figure}
\includegraphics[width=8cm]{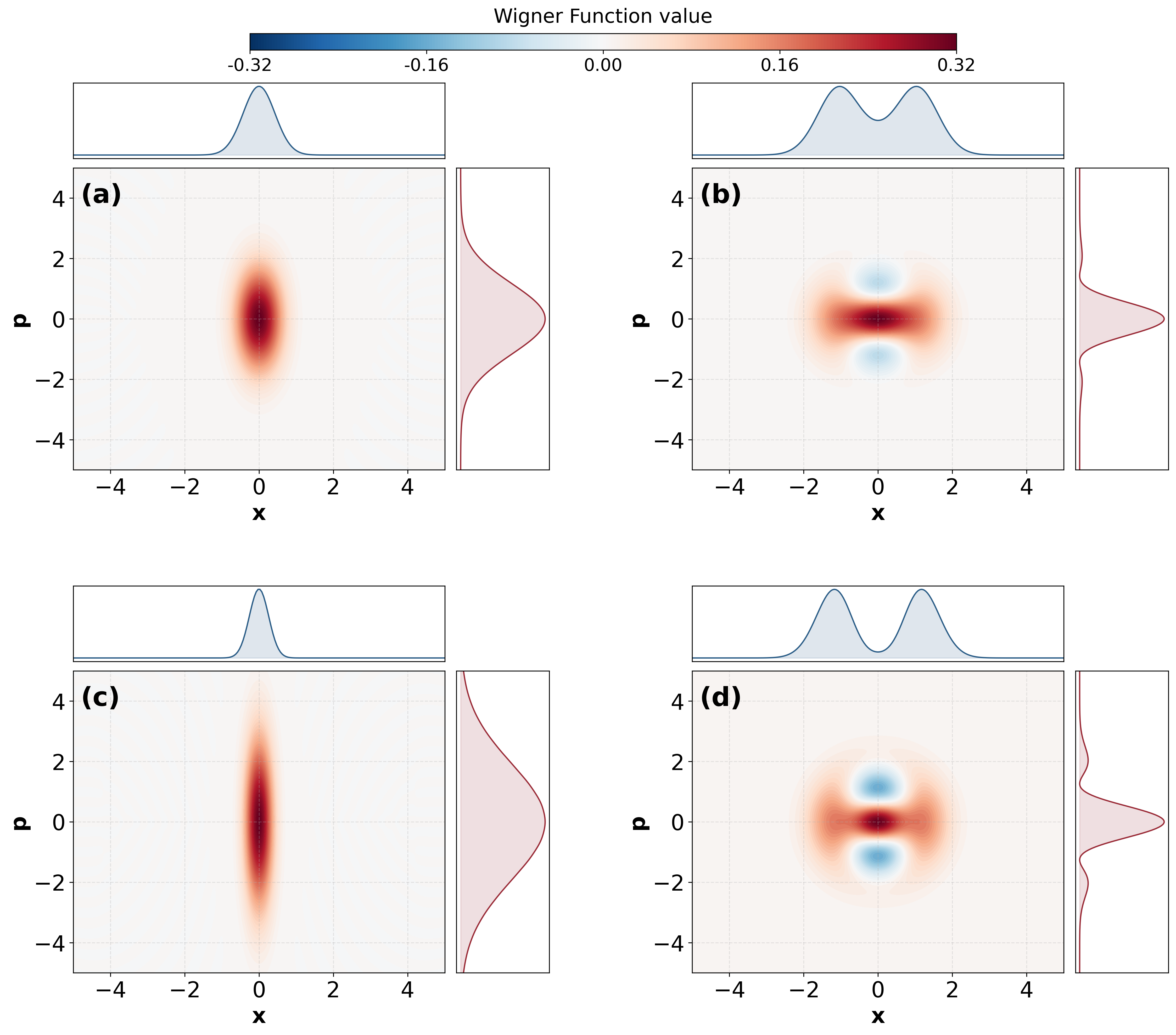}

\caption{\label{fig:2}Wigner functions of the input and output states. (a)
and (c) show the Wigner functions of the input SV states with squeezing
parameters $r=0.5$ and $r=1.0$, respectively. The corresponding
Wigner functions of the output signal states $\vert\Psi^{\prime}\rangle$
are shown in (b) and (d). The amplitude gain is fixed at $g=2.5$. }
\end{figure}

The physical mechanism enabling the effective $a^{2}$ operation can
be understood as follows. Our protocol implements an effective two-photon
subtraction ($a^{2}$) via a heralded, entangled measurement within
a single OPA. The detection of a single idler photon projects the
signal mode into the state $\vert\Psi^{\prime}\rangle$ in Eq. (\ref{eq:26}).
Its form, $S(\xi^{\prime\prime})\left[c_{0}\vert0\rangle+c_{2}\vert2\rangle\right]$,
is mathematically equivalent to the result of applying $a^{2}$ to
a SV, $a^{2}\vert\eta\rangle\propto S(r)\left[\vert0\rangle-\sqrt{2}\tanh r\vert2\rangle\right]$
(see Eqs. (\ref{eq:13}) and (\ref{eq:eq34})). Although other operations
(e.g., $a^{\dagger2},aa^{\dagger},a^{\dagger}a$) can in principle
produce states with a similar two-component structure, the heralding
process in our protocol actively selects for the specific coefficient
ratio $c_{2}/c_{0}$. This ratio is tunable through the OPA gain $g$
and input squeezing parameter $r$, allowing it to be matched precisely
to the target ratio $-\sqrt{2}\tanh r$, which is the characteristic
signature of an ideal $a^{2}$ operation. In this way, the scheme
achieves a fully integrated and heralded realization of two-photon
subtraction, thereby replacing the cascaded discrete optical components
of conventional setups with a single nonlinear element.

To visualize the nG character of the output state, we numerically
compute its Wigner function. As shown in Fig. \ref{fig:2}, the Wigner
functions of the input and output states are compared to illustrate
the effects of the considered process. For the input signal states,
panels $(a)$ and $(c)$ display the characteristic Gaussian profiles
of SV states with squeezing parameters $r=0.5$ and $r=1.0$, respectively.
The increased squeezing from $(a)$ to $(c)$ is evident as the Wigner
function becomes more elongated along one quadrature. The corresponding
output states, shown in panels $(b)$ and $(d)$ for a gain of $g=2.5$,
exhibit significant distortion and the emergence of pronounced negative
regions (blue areas) in the phase space. These negative values are
a direct signature of nonclassical properties, such as quantum interference,
inherent to the output states. The clear structural change from a
Gaussian input to a nG output underscores the strongly nonclassical
nature of the resulting state.

\begin{figure}
\includegraphics[width=8cm]{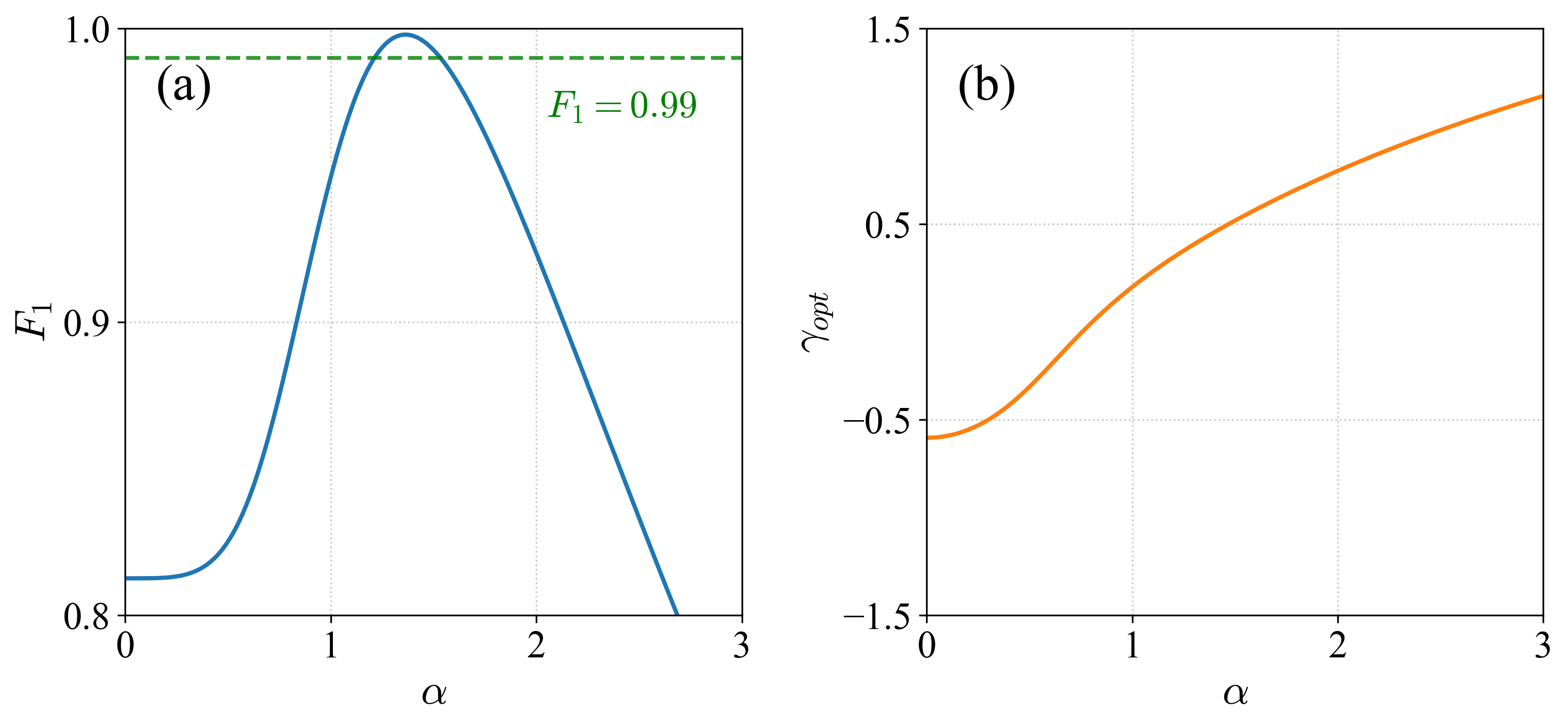}

\caption{\label{fig:3} (a) Optimal fidelity between output state $\vert\Psi^{\prime}\rangle$
and the target squeezed even SC state. (b) Corresponding squeezing
parameter $\gamma$ of the target squeezed SC state that maximizes
the fidelity in panel (a). The squeezing parameter of the initial
SV state is fixed at $r=1.0$. Other parameters are the same as in
Fig. \ref{fig:2}. }

\end{figure}

As established in previous works \citep{PhysRevA.78.063811}, a squeezed
superposition of the vacuum and two-photon Fock states can provide
an excellent approximation to a large-amplitude squeezed even SC state,
defined as
\begin{equation}
\vert\Psi_{sscs}\rangle=\mathcal{N}_{sscs}S(\gamma)\left[\vert\alpha\rangle+\vert-\alpha\rangle\right],\label{eq:27-1}
\end{equation}
where the normalization constant is $\mathcal{N}_{sscs}=(2+2e^{-2\alpha^{2}})^{-1/2}$.
To quantify this, we numerically optimize the parameters $\gamma$
and real amplitude $\alpha$ to maximize the fidelity $F=\vert\langle\Psi_{sscs}\vert\Psi^{\prime}\rangle\vert^{2}$
between the heralded state $\vert\Psi^{\prime}\rangle$ and an ideal
squeezed even SC state. The fidelity is given by 
\begin{align}
F_{1} & =\frac{4\nu e^{\frac{-2\alpha^{2}}{1+\nu^{2}}}}{1+\nu^{2}}\left|\chi\mathcal{N}_{sscs}\left(\sqrt{2}+\mu^{*}\frac{1-\nu^{4}+4\nu^{2}\alpha^{2}}{(1+\nu^{2})^{2}}\right)\right|^{2},\label{eq:43}
\end{align}
where $\nu=e^{-(r^{\prime\prime}-\gamma)}$, $\mu=\frac{c_{2}}{c_{0}}$,
and $\chi=(1+\vert\mu\vert^{2})^{-\frac{1}{2}}$. In Fig. \ref{fig:3},
we present the optimal fidelity between $\vert\Psi^{\prime}\rangle$
and $\vert\Psi_{sscs}\rangle$ as a function of the coherent amplitude
$\alpha$ for fixed squeezing parameter $r^{\prime\prime}=1.0$. For
each value of $\alpha$ in this plot, the squeezing parameter $\gamma$
of the target state is optimized to maximize the fidelity with the
generated state $\vert\Psi^{\prime}\rangle$. As shown in Fig. \ref{fig:3}
for a fixed $r=1.0$, the optimal fidelity reaches $F_{1}=0.998$
at $\alpha\approx1.38$, and $\gamma=0.45$, indicating an extremely
high degree of similarity between the generated state and the ideal
squeezed even SC state. Previous studies have typically prepared the
SC states using conditional heralding schemes based on photon subtraction
from squeezed light \citep{PhysRevA.55.3184,2006,PhysRevLett.97.083604,Wakui:07,PhysRevA.103.013710}.
A $n$- photon subtracted squeezed state $a^{n}S(\gamma)\vert0\rangle$,
is a good approximation to certain squeezed SC states. For large $n$,
the achievable amplitude of the SC state is at most $\alpha=\sqrt{n}$
\citep{PhysRevA.78.063811}. The numerical data presented in Table
\ref{tab:1} corroborate this finding. 

\begin{table}[h]
\raggedright
\caption{\label{tab:1}Optimized parameters for the fidelity between the generated
state$\vert\Psi^{\prime}\rangle$ and an ideal squeezed even SC state.}

\begin{tabular*}{8cm}{@{\extracolsep{\fill}}cccc}
\toprule 
$\boldsymbol{g}$ & $\boldsymbol{\gamma}$ & $\boldsymbol{\alpha}$ & $\boldsymbol{F_{1}}$\tabularnewline
\midrule 
1.0 & 1.0 & 0 & 1\tabularnewline
1.5 & 0.6158 & 1.2245 & 0.999\tabularnewline
2.5 & 0.4470 & 1.3796 & 0.998\tabularnewline
5.0 & 0.3689 & 1.4184 & 0.997\tabularnewline
7.5 & 0.3500 & 1.4184 & 0.997\tabularnewline
10.0 & 0.3440 & 1.4184 & 0.997\tabularnewline
\bottomrule
\end{tabular*}

\end{table}

Table \ref{tab:1} summarizes the optimized parameters and the corresponding
fidelity between the output state generated under a fixed input SV
state with $r=1.0$---and a target squeezed even SC state, evaluated
over a range of gain values $g$. When $g=1.0$, the output state
is identical to the input SV state. As $g$ increases from $1.5$
to $10.0$, the optimal value of $\gamma$ decreases, while the optimal
$\alpha$ increases, saturating at approximately $1.4184$ for $g\geq5.0$.
Notably, the fidelity remains high ($F_{1}\geq0.997$) across the
entire range of $g$, demonstrating the robustness and effectiveness
of the protocol for generating the target nG state with extremely
high fidelity. These results suggest that for a fixed $r$, the generated
state converges to a squeezed even SC state with well-defined coherent
amplitude, which becomes independent of $g$ for sufficiently large
gains. In other words, for an SV signal input, our protocol generates
a squeezed even SC state $\vert\Psi_{sscs}\rangle$ with coherent
amplitude $\alpha\approxeq\sqrt{2}$ and high fidelity. In this sense,
our scheme with an SV input is equivalent to two-photon subtraction,
$a^{2}\vert\eta\rangle$, from the SV state \citep{PhysRevA.78.063811,PhysRevA.77.062315},
achieving nearly perfect fidelity ($F=0.999$). A key advantage of
our scheme is that it generates the squeezed even SC state without
explicitly performing two-photon subtraction.

Recent experimental works have demonstrated the generation of squeezed
odd SC states with small ($\alpha<1$) and larger ($\alpha\approx1.4$)
amplitudes using a continuous-wave low-loss waveguide OPA \citep{Takase:22}
and a conventional OPA \citep{20226}, respectively. In both cases,
an SV state is first prepared using an OPA. Subsequently, Ref. \citep{Takase:22}
employs the traditional scheme of a single-photon subtraction via
an unbalanced beam splitter to generate a squeezed odd SC state with
a fidelity of $F\approx0.55$. In contrast, Ref. \citep{20226} uses
a second OPA to generate a larger-amplitude squeezed odd SC state
with a fidelity of $F\approx0.61\backsim0.65$. The key distinction
of our model is that the OPA is not used directly as a source of squeezed
light. Instead, it functions analogously to a beam splitter, primarily
generating entanglement between the signal and idler modes. Furthermore,
for an SV input, the heralding of a single photon in the idler output
mode causes our scheme to effectively perform a continuous two-photon
subtraction operation. Consequently, the output state is not a squeezed
odd SC state, as in conventional schemes, but rather a squeezed even
SC state with a larger coherent amplitude. Therefore, our model constitutes
a new and feasible scheme for generating squeezed SC states with larger
amplitudes and extremely high fidelity, outperforming traditional
photon-subtraction methods.

To further quantify the overall extent of negativity in the Wigner
distributions with changing the amplitude gain $g$, we use the Wigner-negative
volume \citep{Kenfack2004} . The Wigner-negative volume is defined
as:
\begin{equation}
N=\frac{1}{2}\int\int\left[\vert W(x,y)\vert-W(x,y)\right]dxdy,\label{eq:27}
\end{equation}
 where $W(x,y)$ is the Wigner function of a given state. 

\begin{figure}
\includegraphics[width=8cm]{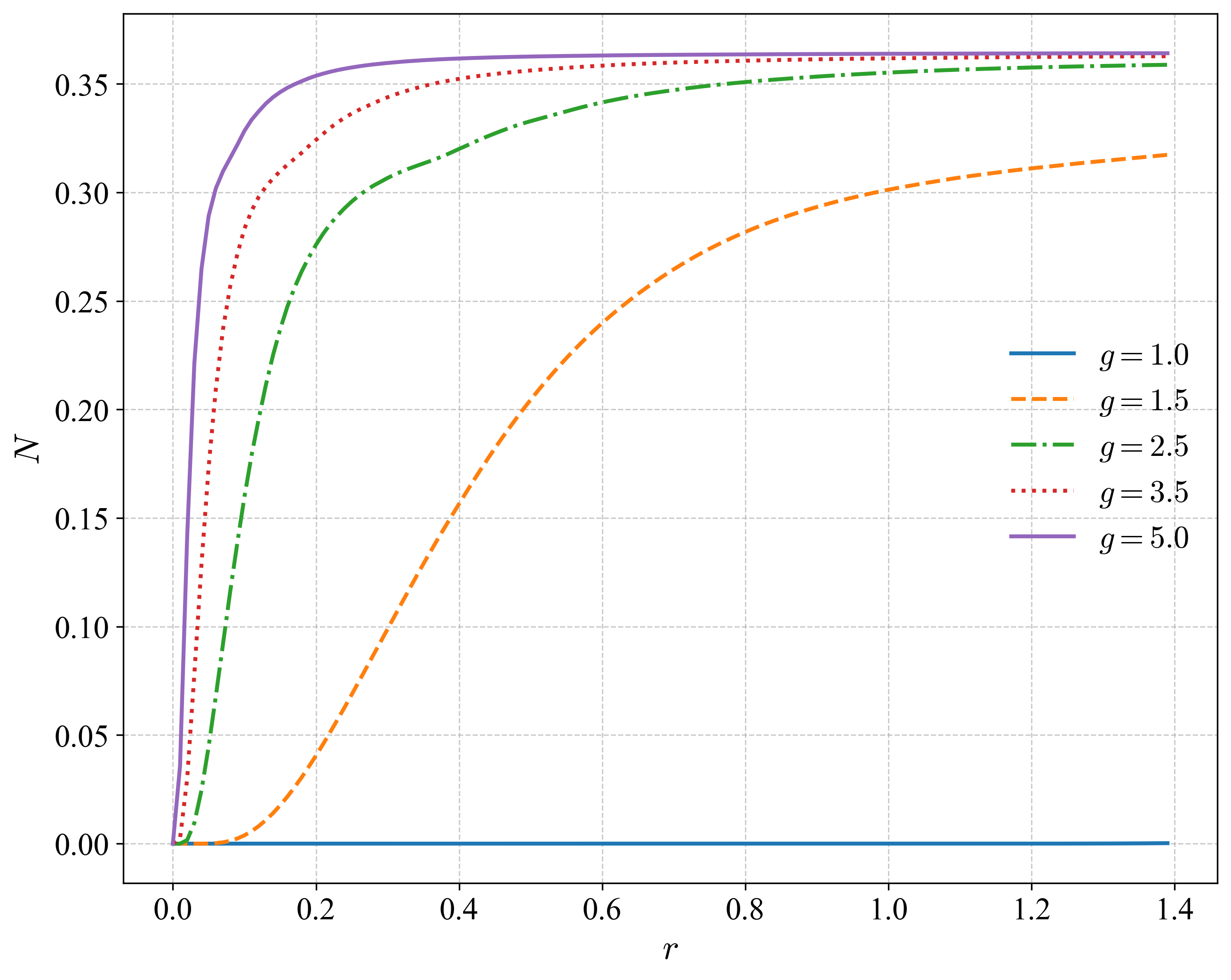}

\caption{\label{fig:4}Dependence of the Wigner-negative volume $N$ \textcolor{blue}{of
the output state $\vert\Psi^{\prime}\rangle$} on the squeezing parameter
$r$ for different values of the amplitude gain $g$.}

\end{figure}

Figure \ref{fig:4} shows the Wigner-negative volume $N$ of the output
state $\vert\Psi^{\prime}\rangle$ as a function of the squeezing
parameter $r$ for different values of the amplitude gain $g$. For
$g=1.0$ (blue solid line), the negativity volume $N$ remains nearly
zero for all $r$, consistent with the input SV state, which has a
positive Wigner function. As $g$ increases, Wigner negativity emerges
and becomes more pronounced. For any $g>1$, the negativity volume
$N$ increases monotonically with the squeezing parameter $r$. Furthermore,
for a fixed squeezing parameter $r$, a larger gain $g$ yields a
higher value of $N$, signifying a stronger nonclassical character
in the generated state. The saturation of the curves at larger $r$
indicates a limit on the achievable negativity volume for a given
$g$. The results in Table \ref{tab:1} and Fig. \ref{fig:4} demonstrate
that the output state asymptotically approaches an even squeezed SC
state with fixed parameters as the gain $g$ and initial squeezing
parameter $r$ increase.

\section{\label{sec:4}Small-Amplitude SC State Signal Input }

We now consider the case where the signal input is a small-amplitude
SC state. These states are defined as 
\begin{equation}
\vert\phi\rangle_{\alpha}=\vert SCS_{\pm}(\alpha)\rangle=\mathcal{N}_{\alpha\pm}\left(\vert\alpha\rangle\pm\vert-\alpha\rangle\right),\label{eq:25}
\end{equation}
where the normalization constants are given by $\mathcal{N}_{\alpha,\pm}=(2\pm2e^{-2\alpha^{2}})^{-1/2}$.
SC states can be generated via several approaches, including photon
subtraction from squeezed vacuum states \citep{2006} and cavity QED
or circuit QED schemes \citep{PhysRevA.93.033853,PhysRevA.106.043721}.
These techniques enable the preparation of optical and microwave SC
states with controllable amplitudes, making them ideal testbeds for
investigating nG quantum processes.

For the SC signal input states, the corresponding normalized output
signal states is given by 
\begin{align}
\vert\Phi\rangle_{\pm} & =\mathcal{N_{\pm}}\left[\frac{1}{g^{2}}\vert\phi\rangle_{\alpha/g}-G^{2}a^{\dagger}a\vert\phi\rangle_{\alpha/g}\right],\nonumber \\
 & =\mathcal{N_{\pm}^{\prime}}\left[\frac{1}{g^{2}}\left(\vert\frac{\text{\ensuremath{\alpha}}}{g}\rangle\pm\vert-\frac{\text{\ensuremath{\alpha}}}{g}\rangle\right)-\frac{G^{2}\text{\ensuremath{\alpha}}}{g}a^{\dagger}\left(\vert\frac{\text{\ensuremath{\alpha}}}{g}\rangle\mp\vert-\frac{\text{\ensuremath{\alpha}}}{g}\rangle\right)\right],\label{eq:31}
\end{align}
where the normalization constant $\mathcal{N_{\pm}^{\prime}}=\left[2(\beta^{2}+G^{4}\vert\frac{\alpha}{g}\vert^{2})\right]^{-1/2}$
with $\beta=\frac{1}{g^{2}}-G^{2}\vert\frac{\alpha}{g}\vert^{2}$.
The output state is a superposition of the initial SC state and a
photon-subtracted-and-added SC state. For larger amplitude gain $g$
the second term which proportional to $a^{\dagger}a$ becomes dominant.
While the photon number operator $a^{\dagger}a$ typically enlarges
the state, the output amplitude in our protocol is attenuated by a
factor of $\frac{\alpha}{g}$. This trade-off prevents amplitude enlargement,
allowing the operation to primarily alter the state's nonclassicality.
The single-trial success probability for this process is given by
$\eta e^{-\vert G\alpha\vert^{2}}|\mathcal{N_{\pm}}|^{-2}$. For the
parameter ranges considered in this work---with the gain $g$ between
$1$ to $5$ and the parameter $\alpha$ between $0$ to $1.5$---this
probability lies between $10^{-3}$ to $10^{-1}$. More specifically,
the probability to herald the even SC input state $\vert\Phi\rangle_{+}$
is on the order of $10^{-3}$ per trial, while for an odd SC input
state, the corresponding probability is approximately $10^{-2}$.

If the gain is set to a critical value, $g=g_{0}=\frac{1}{\sqrt{1-1/\vert\alpha\vert^{2}}},$
the output state simplifies to the following unnormalized form:
\begin{align}
\vert\Phi^{\prime}\rangle_{\pm} & =-\frac{1}{\sqrt{2}}\left[D\left(\frac{\alpha}{g_{0}}\right)\mp D\left(-\frac{\alpha}{g_{0}}\right)\right]\vert1\rangle\nonumber \\
 & =\frac{g}{\sqrt{2}\alpha}\mathcal{N}_{\alpha/g_{0},\pm}^{-1}\left(\frac{\vert\alpha\vert^{2}}{g_{0}^{2}}-a^{\dag}a\right)\vert\phi\rangle_{\alpha/g_{0}}.\label{eq:38}
\end{align}
 In this case, the output corresponds to a superposition of displaced
number states. 

\begin{figure}
\includegraphics[width=8cm]{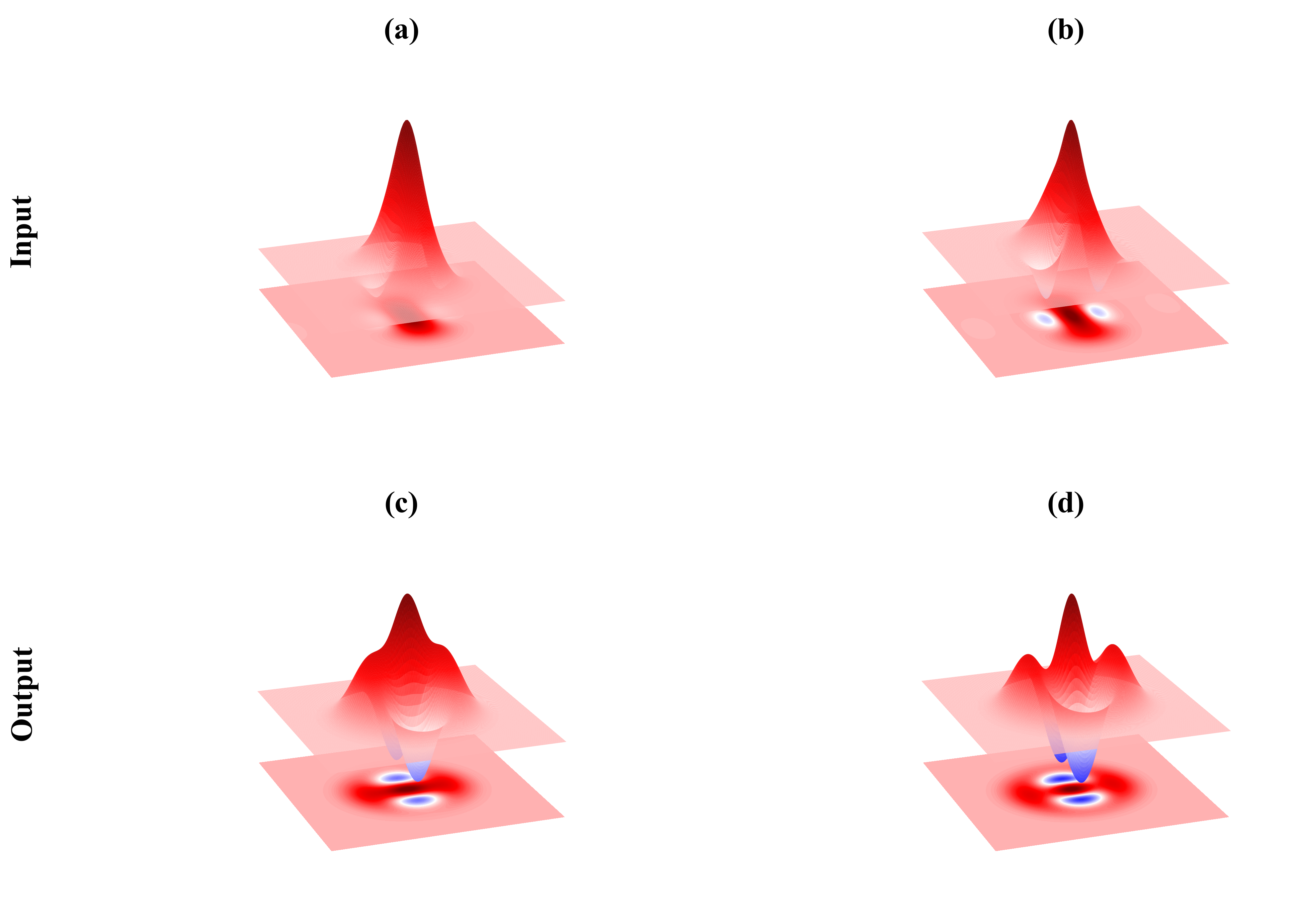}

\caption{\label{fig:5}(a) and (b) show the Wigner functions of the input even
SC state with coherent amplitudes $\alpha=0.8$ and $\alpha=1.01$,
respectively. The corresponding Wigner functions of the output signal
states are shown in (b) and (d), generated with amplitude gains $g=5$
and $g=g_{0}$, respectively.}
\end{figure}

Figure \ref{fig:5} shows the Wigner function distributions of the
input even SC states and their corresponding output states. Panels
$(a)$ and $(b)$ present the Wigner functions of the input even SC
states with coherent amplitudes $\alpha=0.8$ and $\alpha=1.01$,
respectively. Due to the relatively small coherent amplitudes, both
input states exhibit nearly Gaussian profiles, lacking pronounced
nG features or significant negative regions. In contrast, panel $(c)$
shows the output state for the input in $(a)$ with a gain $g=5$.
A distinct negative region emerges, indicating a considerable enhancement
of non-Gaussianity. This trend is more pronounced in panel $(d)$,
which shows the output for input from $(b)$ at the critical gain
$g=g_{0}$. The state exhibits broader and deeper negative regions,
confirming the effective amplification of non-Gaussianity. Furthermore,
for the $g=g_{0}$ and $\alpha=1.01$ {[}see Fig. \ref{fig:5} (d){]},
the generated state has an extremely high fidelity ($F_{2}>0.999$)
with the state $\vert\phi_{2}\rangle=\vert0\rangle-1.416\vert2\rangle$. 

The origin of this specific state can be explained as follows. For
$g=g_{0}$ and $\alpha=1.01$ (giving $\alpha/g_{0}=0.142$), the
output state in Eq. (\ref{eq:27}) becomes $0.02\vert\phi\rangle_{0.142}-0.98a^{\dagger}a\vert\phi\rangle_{0.142}$
(unnormalized), where $\vert\phi\rangle_{0.142}=\vert SCS_{+}(0.142)\rangle$
is the even SC state with amplitude $\alpha=0.142$. An even SC state
with $\alpha\apprle0.75$ closely resembles ($F>0.99$) a Gaussian
SV state \citep{PhysRevA.78.052304}. Consequently, the output state
is equivalent to $0.02S(r_{opt})\vert0\rangle-0.98a^{\dagger}aS(r_{opt})\vert0\rangle$
where $r_{opt}$ is the squeezing parameter that optimizes the fidelity
and depends on $\alpha$. Using the relations in Eqs. (\ref{eq:13}-\ref{eq:36})
and choosing an optimized $r_{opt}$ ($\approx0$) yields the state
$\vert\phi_{2}\rangle$. 

\begin{figure}
\includegraphics[width=8cm]{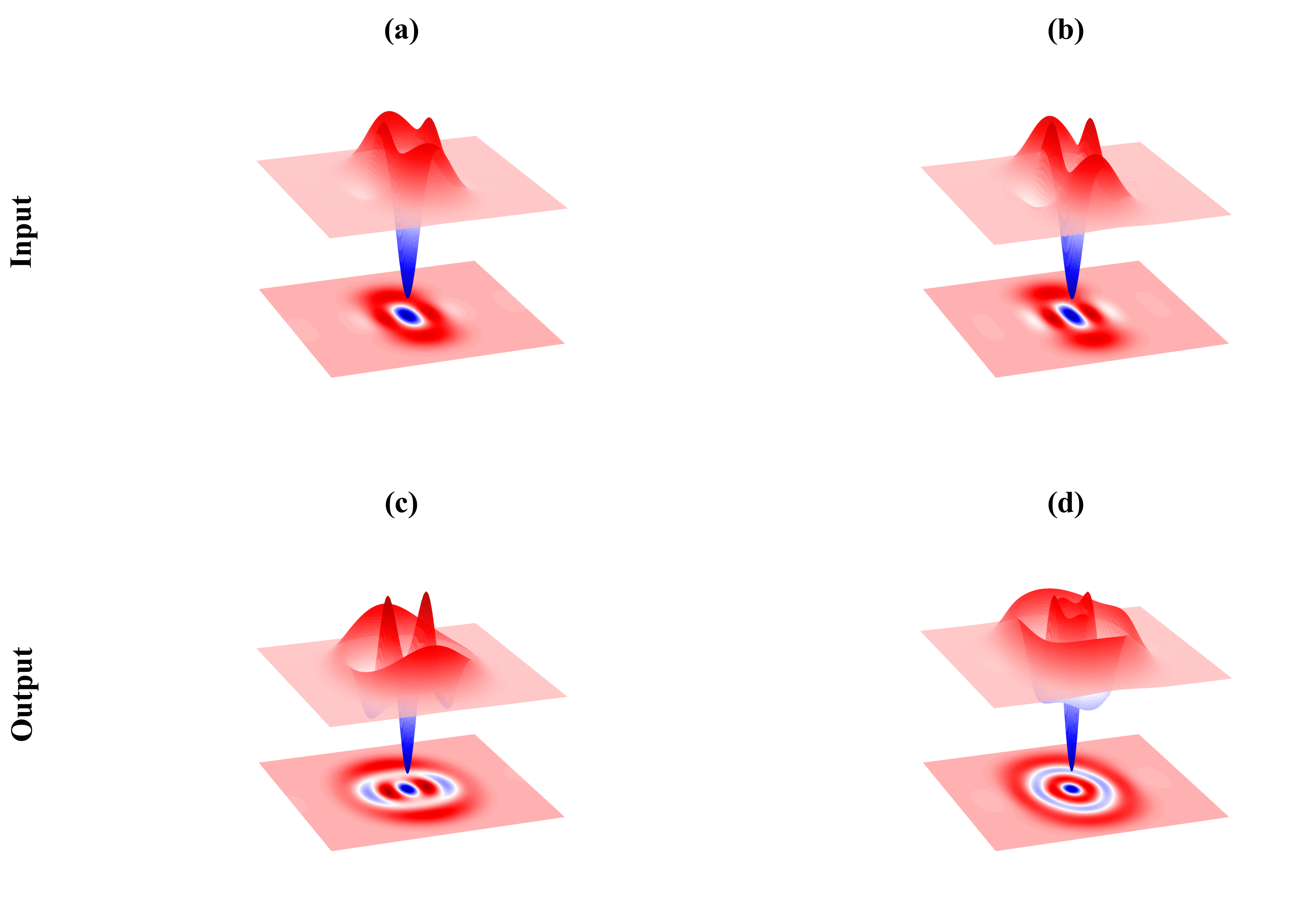}

\caption{\label{fig:6}(a) and (b) show the Wigner functions of the input odd
SC state with coherent amplitudes $\alpha=1.2$ and $\alpha=1.5$,
respectively. The corresponding Wigner functions of the output signal
states are shown in (c) and (d), obtained with amplitude gains of
$g=1.5$ and $g=g_{0}$, respectively.}
\end{figure}

Figure \ref{fig:6} illustrates the evolution of the Wigner functions
when an odd SC state with small coherent amplitude is used as the
input. Panels $(a)$ and $(b)$ show the Wigner functions of the input
odd SC states with coherent amplitudes $\alpha=1.2$ and $\alpha=1.5$,
respectively. These states are characterized by a deep negative dip
at the origin of phase space, reflecting the odd parity of their wavefunctions.
The corresponding output states, obtained with gains $g=1.5$ panel
$(c)$ and $g=g_{0}$ panel $(d)$, exhibit pronounced structural
changes. In particular, the output state in panel $(d)$, generated
at the critical gain, develops intricate ring-like patterns and closely
approximates a squeezed three-photon number state, $\frac{1}{\sqrt{6}}S(r)a^{\dagger3}\vert0\rangle$,
with a fidelity exceeding $F_{3}=0.99$ for an optimal squeezing parameter
$r=-0.23$. The emergence of this squeezed three-photon state can
be explained as follows. In the small-amplitude regime ($\alpha\le1.2$),
an odd SC state is well approximated by a squeezed single-photon state,
$S(r_{opt}^{\prime})\vert1\rangle=\frac{1}{\cosh r_{opt}^{\prime}}a^{\dagger}S(r_{opt}^{\prime})\vert0\rangle,$
with a fidelity exceeding $0.99$ \citep{2006}. The odd SC state
with amplitude $\alpha/g_{0}=1.12$ in Eq. (\ref{eq:27}) can be approximated
by $a^{\dagger}S(r_{opt}^{\prime})\vert0\rangle$. Subsequent calculation
shows that the output then approximates $\vert\phi_{3}\rangle=\frac{1}{\sqrt{6}}S(r_{opt}^{\prime})a^{\dagger3}\vert0\rangle$,
with $r_{opt}^{\prime}=-0.23$. 

Conventional methods for generating photon-number states typically
involve producing a TMSV state from a nonlinear crystal (e.g., BBO)
and performing photon-number-resolving detection (PNRD) on the idler
mode using a visible-light photon counter (VLPC) \citep{Waks_2006}.
A successful detection of $n$ photons in the idler arm heralds the
preparation of the photon-number state $\vert n\rangle$ in the signal
mode. However, such schemes impose stringent requirements on the detector.
While current PNRDs can reliably resolve large photon numbers \citep{RN17},
they are nevertheless limited by a significant drop in overall detection
efficiency as $n$ increases \citep{erkilic2025unifiedopticalplatformnongaussian}.
As demonstrated in recent experimental work \citep{RN16}, the heralding
probability for adding three photons to a small-amplitude coherent
state is extremely low, merely on the order of $10^{-9}$ per pulse.
In contrast, our scheme---using an odd SC state as an input---requires
the detection of only a single photon in the idler mode to herald
the generation of an approximate three-photon Fock state in the signal
mode. The numerical analysis showed that the successful heralding
probability of our state $\vert\phi_{3}\rangle$ is on the order of
$10^{-2}$ with high fidelity ($F_{3}>0.99$). Thus, our protocol
provides a resource-efficient method for generating specific photon-number
states, circumventing the need for high-performance PNRDs. This approach
is readily scalable and adaptable to a broad range of quantum optical
platforms. Although the resulting state exhibits a small degree of
squeezing, this can be compensated for with an additional OPA stage.
Consequently, the resulting (squeezed) three-photon state has potential
applications in quantum error correction \citep{PhysRevA.55.900,2024Q},
quantum channel capacity optimization \citep{RevModPhys.58.1001,RevModPhys.66.481},
and quantum phase estimation \citep{2023N,e27070712}. In previous
work \citep{20226}, sequential usage of OPA implements a single-mode
squeezing operation to generate small amplitude SC states heralded
with single-photon subtraction with the aim of first using an additional
beam splitter and then amplifying the state. In contrast, our heralding
scheme employs the OPA differently; its overall effect is not to amplify
the initial SC state but to significantly enhance its non-Gaussianity.

\begin{figure}
\includegraphics[width=8cm]{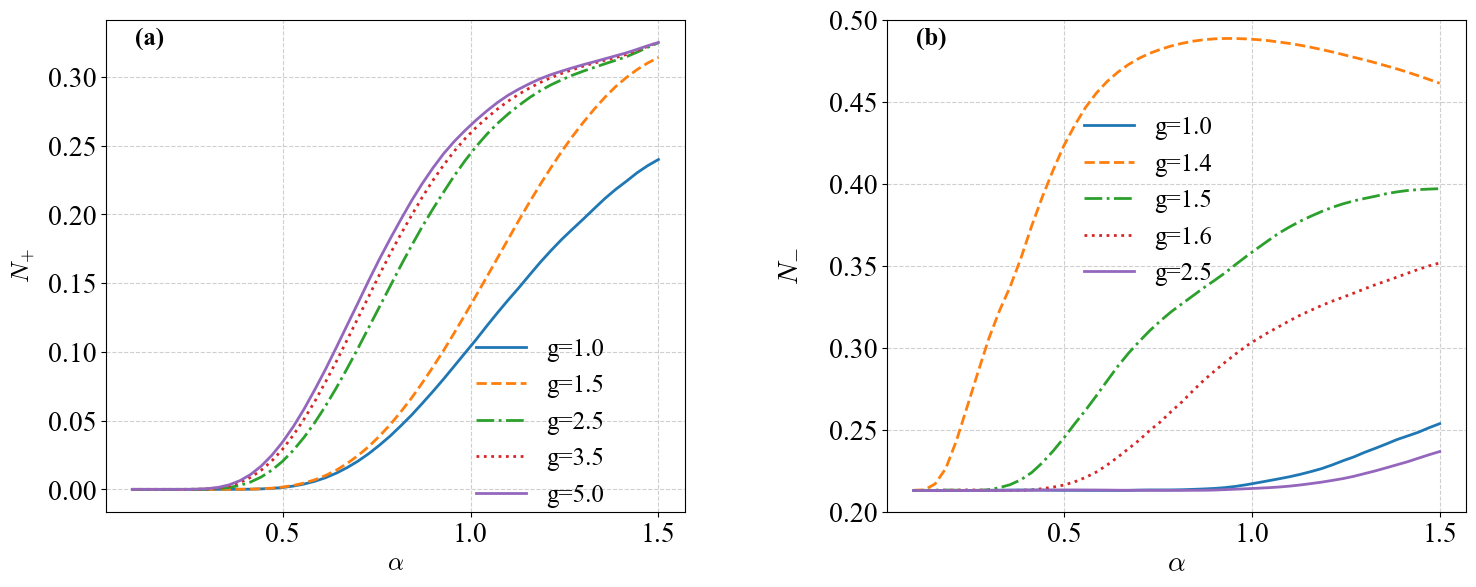}

\caption{\label{fig:7} Dependence of the Wigner-negative volume $N$ \textcolor{blue}{of
output state $\vert\Phi^{\prime}\rangle_{\pm}$} on the coherent state
amplitude $\alpha$ for different values of the amplitude gain $g$. }
\end{figure}

To further investigate the effect of the OPA on input SC states, we
plot the Wigner-negative volume $N$ as a function of state amplitude
$\alpha$ for different amplitude gains $g$, as shown in Fig. \ref{fig:7}.
Figure \ref{fig:7} (a) shows the results for an even SC state input.
The negativity $N_{+}$ increases monotonically with $\alpha$ for
all amplitude gains $g$, eventually saturating at larger amplitudes.
For all $g>1$, the value of $N_{+}$ is greater than that of the
initial state ($g=1$), demonstrating the effectiveness of the OPA
in enhancing the non-Gaussianity of the even SC state. In contrast,
Fig. \ref{fig:7} (b) presents the case of an odd SC state input,
where $N_{-}$ exhibits a non-monotonic dependence on $\alpha$ for
different amplitude gains $g$. The negativity first reaches a maximum
for an intermediate gain (e.g., $g=1.4$) and then decreases as $g$
increases further. 

These behaviors can be understood from the form of the output state
in Eq. (\ref{eq:31}). For an even SC input, the nG character of the
output state is dominated by the term $\frac{G^{2}\text{\ensuremath{\alpha}}}{g}a^{\dagger}\left(\vert\frac{\text{\ensuremath{\alpha}}}{g}\rangle-\vert-\frac{\text{\ensuremath{\alpha}}}{g}\rangle\right)$,
whose relative weight increases with $g$. Since this component resembles
an odd SC state, it retains significant nonclassicality even in the
limit $\frac{\text{\ensuremath{\alpha}}}{g}\rightarrow0$. Consequently,
$N_{+}$ grows with $\alpha$ and eventually saturates. In contrast,
for an odd SC input, the dominant nG component is $\frac{G^{2}\text{\ensuremath{\alpha}}}{g}a^{\dagger}\left(\vert\frac{\text{\ensuremath{\alpha}}}{g}\rangle+\vert-\frac{\text{\ensuremath{\alpha}}}{g}\rangle\right)$.
However, as $g$ increases and $\frac{\text{\ensuremath{\alpha}}}{g}\rightarrow0$,
the contribution of this term diminishes, leading to a reduction in
nonclassicality. This explains the non-monotonic, rise-and-fall behavior
of $N_{-}$ as a function of $\alpha$ for different amplitude gains
$g$.

In summary, the numerical analysis confirms that the proposed protocol
can effectively manipulate and enhance the non-Gaussianity of both
even and odd SC states.

Our protocol represents a conceptual advance from the original role
of a heralded OPA as a specialized state generator. While the seminal
scheme by Shringarpure and Franson \citep{PhysRevA.100.043802} generated
specific nG states from a coherent input, our device processes arbitrary
non-classical inputs, thereby functioning as a reconfigurable quantum
processor capable of generating a continuous family of nG states.
This versatility is demonstrated by two representative functionalities:
first, when fed with SV, it acts as an integrated two-photon subtractor
to produce squeezed SC states of larger amplitude; second, when driven
by small-amplitude SC states, it serves as a non-Gaussianity amplifier,
distilling high-purity photon-number superpositions.

This integrated approach offers a functional flexibility distinct
from conventional discrete-optics methods---such as sequential probabilistic
photon subtraction from squeezed light, used to generate SC states
\citep{PhysRevA.55.3184,2006,PhysRevA.82.031802,Takase:22,PhysRevA.110.023703}.
Those established schemes, while capable in principle of producing
various states, typically require dedicated optical configurations
optimized for each target state and incur cascaded losses that affect
their overall efficiency and scalability. By contrast, our architecture
integrates reconfigurable nonlinear operations within a single platform,
providing a more versatile route to preparing a broad range of nG
states, with potential advantages in reduced optical loss and simplified
control.

\section{\label{sec:5}influence of photo loss}

\begin{figure}
\includegraphics[width=8cm]{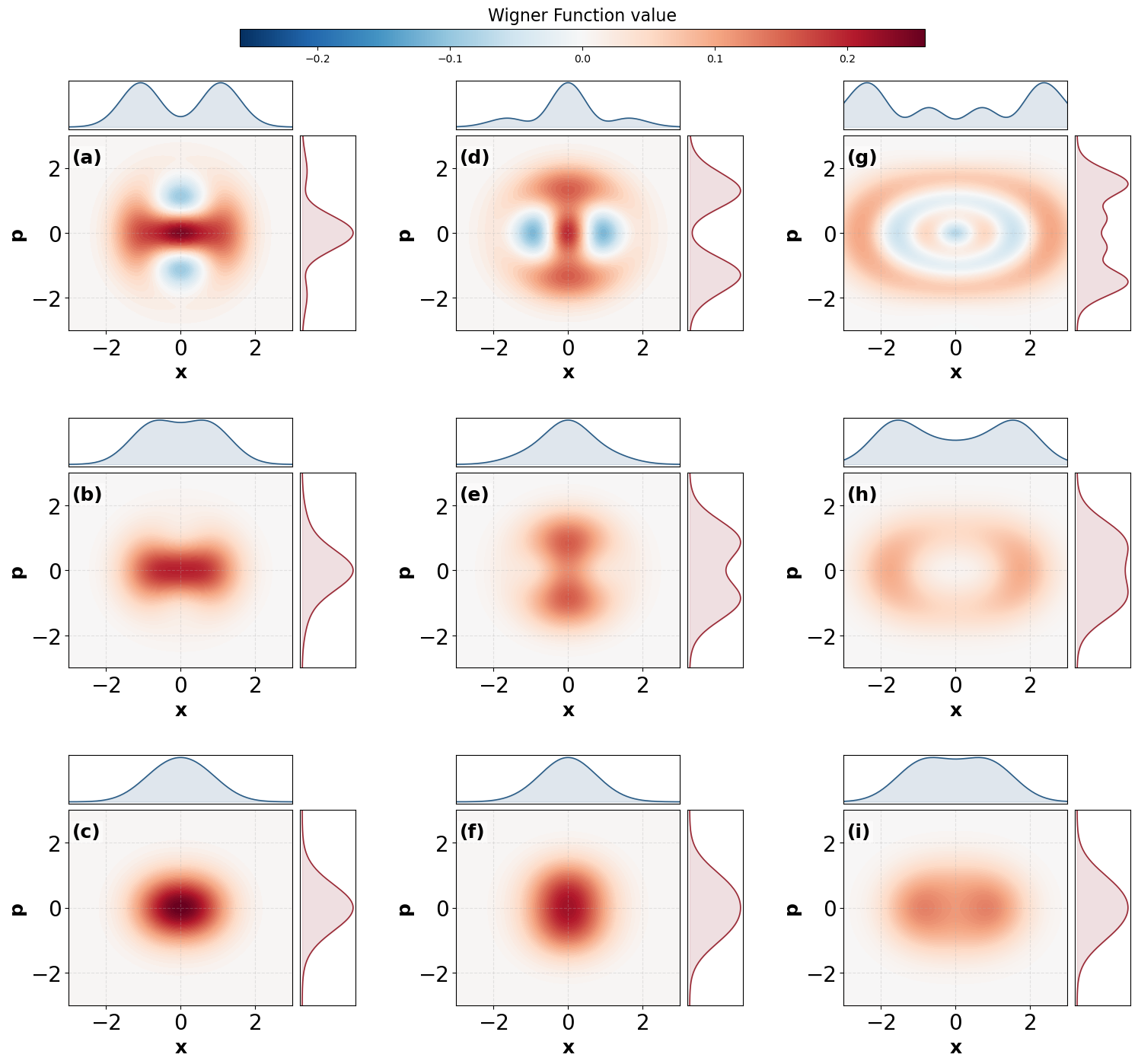}

\caption{\label{fig:8}Wigner functions of the output states under different
dissipation strengths. Panels $(a)$-$(c)$ show the SV input state
with $r=1.0$ and $g=2.5$; $(d)$-$(f)$ display the even SC state
($\alpha=1.2,\,g=g_{0}$); and $(g)$-$(i)$ present the odd SC state
($\alpha=1.5,\,g=g_{0}$). Each row corresponds to different rescaled
time $\kappa t=0.1,\,0.5,\,1.0$, respectively.}
\end{figure}

From a practical perspective, it is essential to analyze the decoherence
of the generated quantum states due to photon loss. When a quantum
state is coupled to an environment, it typically undergoes a transition
from non-Gaussian to Gaussian characteristics. In a photon-loss channel,
the negative regions of the Wigner function---a hallmark of non-Gaussianity---gradually
disappear. The evolution of the density matrix $\rho$, which quantitatively
describes this dynamical behavior, can be modeled using the Lindblad--Gorini--Kossakowski--Sudarshan
(LGKS) master equation \citep{breuer2002theory}:
\begin{equation}
\frac{d\rho}{dt}=\kappa\left(2a\rho a^{\dagger}-a^{\dagger}a\rho-\rho a^{\dagger}a\right),\label{eq:ms}
\end{equation}
where $\kappa$ is the decay rate, and the rescaled time $\kappa t$
characterizes the dissipation strength. The decoherence dynamics are
illustrated in Fig. \ref{fig:8}, which shows the Wigner functions
of the output states obtained by numerically solving the LGKS master
equation for different values of $\kappa t$. The blue regions, representing
negative Wigner function values, gradually shrink and eventually vanish
as $\kappa t$ increases. Furthermore, the separation between the
two peaks in the cat-like states decreases, eventually merging into
a single Gaussian peak. This behavior indicates that the initial nG
states progressively lose their nonclassicality and non-Gaussianity
through interaction with the environment, eventually decaying into
Gaussian states.

\begin{figure}
\includegraphics[width=8cm]{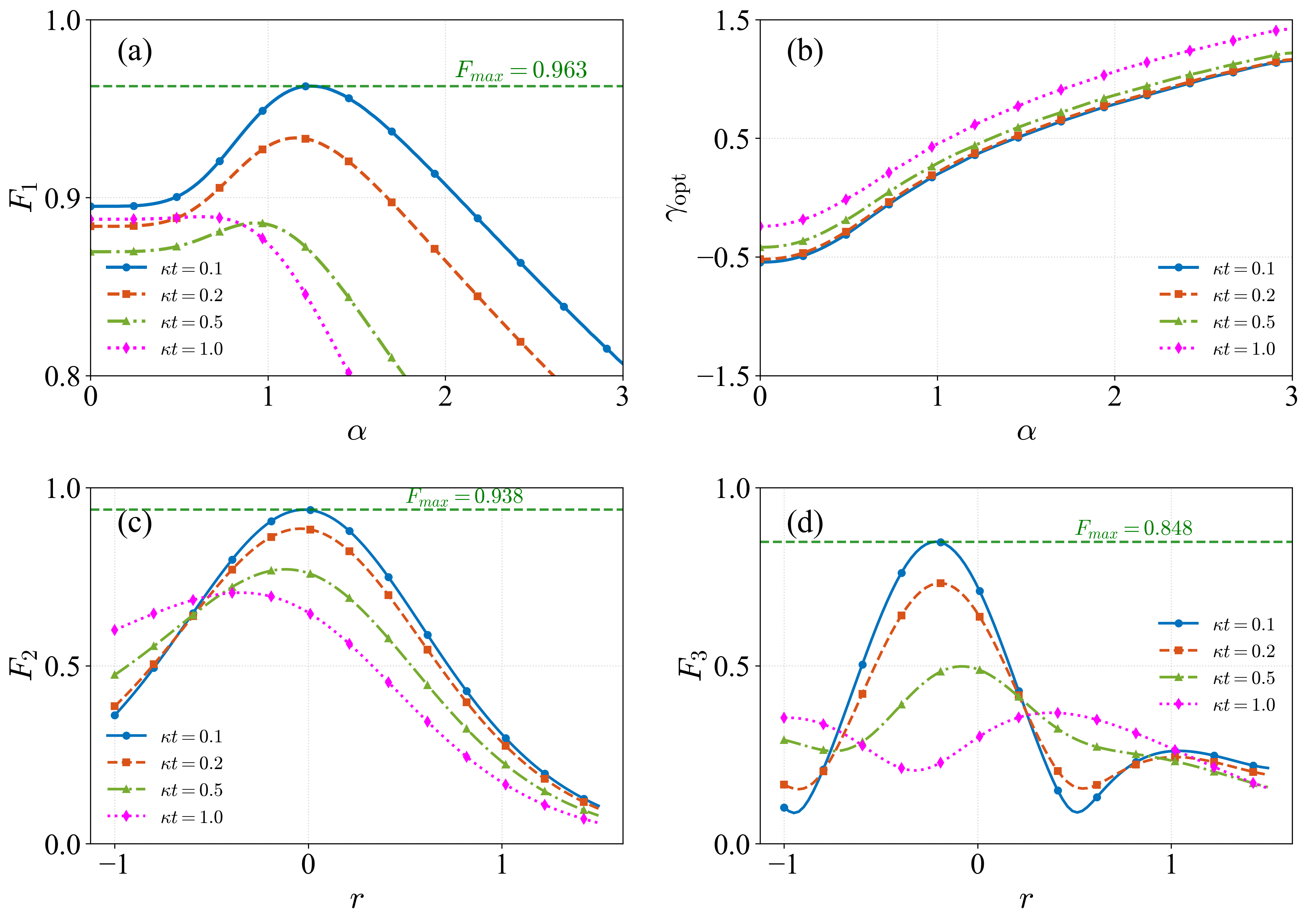}

\caption{\label{fig:9} Fidelity between target states and generated output
states under photon loss. (a) Optimized fidelity between the output
state and the target squeezed even SC state $\vert\Psi_{sscs}\rangle$
as a function of the coherent amplitude $\alpha$ for different photon
dissipation rates $\kappa t$, with fixed parameters $r=1.0$ and
$g=2.5$. (b) Squeezing parameter $\gamma$ of the target squeezed
SC state that maximizes the fidelity shown in panel (a). (c) Fidelity
between the output state and the target state $S(r)(\vert0\rangle-1.416\vert2\rangle)$
as a function of $r$ for an even SC input, with $\alpha=1.01$ and
the gain $g=g_{0}$. (d) Fidelity between the output state and the
target state $S(r)\vert3\rangle$ as a function of $r$ for an odd
SC input, with $\alpha=1.5$ and the gain $g=g_{0}$.}

\end{figure}

To illustrate the effect of photon loss on the quality of the optimized
output states, Figure \ref{fig:9} presents the fidelity between different
target states and the output states generated by our protocol under
photon loss. These output states are obtained by solving the master
equation (see Eq. (\ref{eq:ms})) while accounting for the photon
loss channel.

As shown in Fig. \ref{fig:9}, after passing through a single-photon
loss channel, the optimized fidelity between the target and generated
states decreases monotonically with the dissipation rate. Panels (a)
and (b) show how the optimized fidelity between the target squeezed
even cat state $\vert\Psi_{sscs}\rangle$ and our generated state
changes for the SV input case when photon loss is included. These
results provide a clear comparison with the ideal scenario depicted
in Fig. \ref{fig:3}. For photon loss rates $\kappa t=0.1,\,0.2,$
$0.5$ and 1.0, the maximum fidelity between the output state and
the target squeezed even SC state decreases to $F_{1}=0.96$, $0.93$,
$0.87$ and $0.88$, respectively, while the corresponding coherent
amplitudes are reduced to $\alpha=1.24$, $1.15$, $1.0$, and 0.63. 

However, as shown in panels (c) and (d), for the two SC state input
cases, the maximum fidelity between the loss-affected output states
and the target states is more strongly impacted by dissipation. Taking
$\kappa t=0.2$ as an example, for an SV input, the output state still
maintains a high fidelity of up to 0.93 with the target squeezed SC
state. By contrast, for the two SC state inputs, the corresponding
fidelities drop to 0.88 and 0.73, respectively. This demonstrates
that the protocol with an SV input is more robust against dissipation
than those with SC state inputs---a trend consistent with the gradual
suppression of non-classical features visualized via the Wigner functions
in Fig. \ref{fig:8}.

Notably, for the SV input case, even at a substantial loss level of
$\kappa t=1$, where the Wigner negativity (a stringent witness of
non-classicality) is largely eroded, the overall state fidelity remains
above 0.8 across the entire range of coherent amplitudes considered
($\alpha\in(0,1.418)$; see Fig. \ref{fig:8}). This indicates that
the functional quantum information encoded in these generated states---as
measured by fidelity---degrades more slowly than the visibility of
their fine-grained quantum interference features, as measured by Wigner
negativity. Consequently, under realistically achievable low-loss
conditions ($\kappa t\lesssim0.2$), the output can still be considered
a high-fidelity non-Gaussian resource. This inherent robustness supports
the potential utility of these states in practical quantum information
processing \citep{RN14}.

\section{\label{sec:6}Discussion and conclusion}

\begin{table*}
\centering
\caption{\label{tab:2}Summary of optimized input-output relation under the
effective operations of the OPA-based heralded state generation protocol.}

\begin{tabular*}{16cm}{@{\extracolsep{\fill}}ccc}
\toprule 
\textbf{Input} & \textbf{Output} & \textbf{Effective Operation}\tabularnewline
\midrule
Coherent state \citep{PhysRevA.100.043802}  & SPAC and DSPN states & Integrated $a^{\dagger}$, Non-Gaussianity amplifier\tabularnewline
\midrule
SV state & Squeezed even SC state ($\alpha\approx1.4$)(see Eq. \ref{eq:26}) & Integrated $a^{2}$\tabularnewline
\midrule
Even SC state & Superposition of Fock states ( e.g., $\vert\phi_{2}\rangle$) & Non-Gaussianity amplifier\tabularnewline
\midrule 
Odd SC state & Squeezed three-photon state (e.g., $\vert\phi_{3}\rangle$) & Non-Gaussianity amplifier\tabularnewline
\bottomrule
\end{tabular*}
\end{table*}

In this work, we have generalized the heralded OPA protocol into a
highly versatile and practical source for distinct classes of nG states.
By moving beyond the coherent state inputs of previous scheme, we
have demonstrated that a single, integrated setup can perform two
critical functions: the generation of high-fidelity, larger-amplitude
squeezed SC states from a SV input, and the significant amplification
of non-Gaussianity from small-amplitude SC states, producing tailored
photon number superpositions.

A key finding of our analysis is the protocol's intrinsic equivalence
to an integrated two-photon subtraction for a SV input. This operation
is achieved not with discrete, lossy optical components, but within
the stable environment of the OPA itself, yielding a squeezed even
SC state with a coherent amplitude of $\alpha\approx1.4$---a value
challenging to achieve with sequential single-photon subtraction.
Furthermore, when the input is a small-amplitude SC state, the protocol
acts as a non-Gaussianity engine, distilling states that closely approximate
high-value resources---such as the specific superposition $\vert0\rangle-1.4\vert2\rangle$
and squeezed three-photon state---with fidelities exceeding 0.99.
The robustness of these states is confirmed by their retention of
significant Wigner negativity under initial, realistic levels of photon
loss, underscoring their potential for practical applications. We
provide a summary of the OPA-based heralded state generation protocol
in Table .\ref{tab:2}, which specifies, for a given input quantum
state, its corresponding optimized output state and the effective
operation connecting them. 

The experimental viability of our proposal is firmly supported by
its practical performance metrics. The calculated per-trial success
probabilities, ranging from $10^{-4}$ to $10^{-2}$ for a gain $g\in[1,10]$,
are commensurate with other established heralding schemes including
SPAC states \citep{2004S} and cluster state generation \citep{PhysRevLett.97.110501}.
When combined with modern high-repetition-rate laser sources \citep{Wakui:20,CHEN2025111703},
these probabilities translate to substantial absolute generation rates
of $10^{5}$ to $10^{7}$states per second, confirming the experimental
feasibility of our proposal.

The performance and scalability of this protocol are poised for further
enhancement through technological advances. Progress in low-loss integrated
photonics \citep{PhysRevLett.133.083803} will minimize decoherence,
while advances in high-efficiency superconducting single-photon detectors
\citep{3803,Korzh2020np} will maximize the heralding efficiency and
thus the effective generation rate. The inherent versatility and integrated
nature of our scheme make it highly amenable to these ongoing developments.

In summary, this work significantly expands the toolbox for quantum
state engineering by providing a unified, flexible, and experimentally
accessible platform for generating high-purity nG states. The heralded
OPA protocol thus constitutes a promising and versatile workhorse
for advancing CV quantum computation, metrology, and other related
information processing. 

The key innovation lies in reconceptualizing the heralded OPA from
a state generator into a programmable quantum state processor. This
paradigm shift, supported by the experimental merit of functional
integration within a single nonlinear element, enables a single apparatus
to perform diverse nG operations---from cat-state generation to non-Gaussianity
amplification---determined solely by the input state and gain. Consequently,
this work points to a clear and practical new experimental pathway.
It moves beyond crafting specialized setups for each target state
and could help pave the way toward the development of integrated,
reconfigurable quantum optical circuits. Such circuits would efficiently
bridge between readily available Gaussian resources and the complex
nG states required for fault-tolerant quantum technologies. 
\begin{acknowledgments}
This study was supported by the National Natural Science Foundation
of China (No. 12365005). 
\end{acknowledgments}

\bibliographystyle{apsrev4-1}
\bibliography{Reference}

\end{document}